\documentclass[aps,superscriptaddress,showpacs,notitlepage]{revtex4-1}
\usepackage{epsfig}
\newcommand{\ben}{\begin{eqnarray}}
\newcommand{\een}{\end{eqnarray}}
\newcommand{\nnu}{\nonumber\\}
\newcommand{\bef}{\begin{figure}[htb]\centering}
\newcommand{\eef}{\end{figure}}

\begin{document}
\title{Nuclear modification of high transverse momentum particle production \\
in p+A collisions at RHIC and  LHC}

\date{\today}

\author{Zhong-Bo Kang}
\email{zkang@lanl.gov} 
\affiliation{Theoretical Division, 
                   Los Alamos National Laboratory, 
                   Los Alamos, NM 87545, USA}

\author{Ivan Vitev}
\email{ivitev@lanl.gov}                   
\affiliation{Theoretical Division, 
                   Los Alamos National Laboratory, 
                   Los Alamos, NM 87545, USA}       
                   
\author{Hongxi Xing}
\email{xinghx@iopp.ccnu.edu.cn}
\affiliation{Interdisciplinary Center for Theoretical Study and Department of Modern Physics,
                   University of Science and Technology of China, 
                   Hefei 230026, China}
\affiliation{Institute of Particle Physics, 
                   Central China Normal University,
                   Wuhan 430079, China}

\begin{abstract}
We present results and predictions for the nuclear modification of the differential cross sections for inclusive 
light hadron and prompt photon production in minimum bias d+Au collisions at $\sqrt{s} = 200$~GeV 
and  minimum bias p+Pb collisions at $\sqrt{s} = 5$~TeV at RHIC and LHC, respectively.
Our calculations combine the leading order perturbative QCD formalism with cold nuclear matter
effects that arise from the elastic, inelastic and coherent multiple scattering of partons in large
nuclei. We find that a theoretical approach that includes the isospin effect, Cronin effect, cold nuclear 
matter energy loss and dynamical shadowing  can describe the RHIC d+Au data rather well. 
The LHC p+Pb predictions will soon be confronted by new experimental results to help clarify 
the magnitude and origin of cold nuclear matter effects and facilitate precision dense QCD matter tomography.
\end{abstract}


\maketitle

\section{Introduction}
Medium-induced modification of high transverse momentum ($p_T$) particle production in nucleus collisions (p+A, A+A) 
relative to the naive binary collision-scaled proton-proton (p+p) baseline expectation is a powerful probe of the 
properties of dense QCD matter. In particular, the nuclear modification in d+Au collisions at the RHIC and the 
forthcoming p+Pb results at the LHC can provide valuable information on the elastic, inelastic and coherent multiple 
parton scattering processes inside a large nucleus and are vital testing grounds for novel nontrivial 
QCD dynamics~\cite{Salgado:2011wc}.  The manifestation of such nontrivial QCD dynamics in p+A reactions is usually 
referred to as cold nuclear matter (CNM) effects, as opposed to the effects of parton and particle formation and 
propagation in the ambiance of the quark-gluon plasma created in high energy A+A collisions~\cite{Gyulassy:2003mc}. 

There have been several different approaches to studying  CNM effects. One of them is based on the leading-twist 
perturbative QCD factorization and attributes all these effects to universal nuclear parton distribution 
functions (nPDFs), which are the only ingredients different from the case of p+p 
collisions~\cite{Hirai:2007sx,Eskola:2009uj, deFlorian:2011fp}. Through a global fitting procedure to fix the 
parametrization of nPDFs, this approach can describe  part of the RHIC d+Au data reasonably well and has made 
predictions for  p+Pb collisions~\cite{QuirogaArias:2010wh,Barnafoldi:2011px,Arleo:2011gc}. Another approach is 
the so-called Color Glass Condensate (CGC) 
approach~\cite{McLerran:1993ni,Kharzeev:2003wz,Albacete:2010bs,Tribedy:2011aa,JalilianMarian:2011dt,Albacete:2012xq}. 
It focuses on non-linear corrections to QCD evolution equations in very dense gluonic systems and 
is only applicable in the very small-$x$ region - the so-called classical gluon saturation regime. Various 
predictions for the p+Pb run have also been made based on CGC approach, see  
Refs.~\cite{Tribedy:2011aa,JalilianMarian:2011dt,Albacete:2012xq}. There are also calculations and predictions 
for the nuclear modification factor in Monte Carlo models such as HIJING~\cite{Xu:2012au}.

In this manuscript we follow an approach different from the above mentioned. It is based on perturbative 
QCD factorization and  CNM effects are implemented separately within the formalism. The advantage of 
this approach is that all CNM effects have clear physical origin, mostly centered around the idea of  
multiple parton scattering~\cite{Vitev:2006bi}.  Thus, they can be calculated and their implementations is well documented in the 
literature. They are also process-dependent.  In  this manuscript  we discuss the isospin effect, Cronin effect, 
cold nuclear matter energy loss, and dynamical shadowing. The isospin effect is easily incorporated by 
replacing the nPDFs by the $Z,\, A-Z$-weighted average of proton and neutron PDFs~\cite{Wang:1998ww,Kang:2008wv}. 
Theoretical interpretations of the Cronin effect are reviewed in~\cite{Accardi:2002ik} and our implementation 
is via initial-state parton transverse momentum broadening. Initial-state cold nuclear matter parton energy 
loss was calculated in~\cite{Vitev:2007ve, Xing:2011fb}. Nuclear shadowing can be implemented as coherent power 
corrections to the differential particle production cross section~\cite{Qiu:2004da,Qiu:2003vd}. A perturbative  
QCD formalism with these effects incorporated can be used to understand the qualitative and in most cases 
quantitative features of the nuclear modification in p+A (or d+A) collisions observed at CERN SPS and RHIC 
experiments~\cite{Vitev:2003xu,Vitev:2002pf,Qiu:2004da,Kang:2011bp}. In anticipation of the new LHC p+Pb 
results we present predictions in the framework of this approach to incorporating  CNM effects.  Specifically, 
we focus on inclusive particle production in d+Au and p+Pb collisions  and show results for the nuclear 
modification factor of neutral pions, charged hadrons and prompt photons.

This paper is organized as follows. In Sec. II, we present our approach to studying CNM effects. 
We first review the basic perturbative QCD formalism for single inclusive particle production in 
p+p collisions. We then discuss how various CNM effects are incorporated in this formalism.
In Sec. III, we implement these CNM effects and show that the formalism can describe particle 
production in d+Au collisions at RHIC energies. We then present our theoretical model predictions 
for the upcoming p+Pb collision results at the LHC. We summarize our paper in Sec.~IV.

\section{Single inclusive particle production}
In this section we review the basic perturbative QCD formalism for single inclusive particle production in p+p collisions. 
In particular, we focus on light hadron and prompt photon production. We then discuss the incorporation of various cold nuclear 
matter effects within this formalism. Specifically, we comment on the isospin effect, Cronin effect, cold nuclear matter 
energy loss and dynamical shadowing. 

\subsection{Basic pQCD formalism}

To leading order in the framework of factorized perturbative QCD, single inclusive hadron production in p+p collisions, 
$p(p_1)+p(p_2)\to h(p_h)+X$, can be written as follows~\cite{Owens:1986mp}:
\ben
\frac{d\sigma}{dy d^2p_T} &=& K\frac{\alpha_s^2}{s}\sum_{a,b,c}\int \frac{dx_a}{x_a}d^2k_{aT} \, 
f_{a/N}(x_a, k_{aT}^2)\int \frac{dx_b}{x_b}d^2k_{bT} \, 
f_{b/N}(x_b, k_{bT}^2) 
\nnu
&& \times
\int \frac{dz_c}{z_c^2}\,  D_{h/c}(z_c) H_{ab\to c}(\hat s,\hat t,
\hat u)\delta(\hat s+\hat t+\hat u),
\label{light}
\een
where $y$ and $p_T$ are the rapidity and transverse momentum of the produced hadron and $\sum_{a,b,c}$ runs 
over all parton flavors. In Eq.~(\ref{light})  $s=(p_1+p_2)^2$, $D_{h/c}(z_c)$ is the fragmentation function 
(FF) of parton $c$ into hadron $h$, $H_{ab\to c}(\hat s,\hat t,\hat u)$ are the hard-scattering coefficient 
functions with $\hat s,\hat t,\hat u$ being the usual partonic Mandelstam variables~\cite{Owens:1986mp,Kang:2010vd}. 
We also include a phenomenological $K$-factor to account for higher order QCD 
contributions~\cite{Wang:1998ww,Vitev:2008vk}. $f_{a,b/N}(x, k_{T}^2)$ are the parton distribution functions with 
longitudinal momentum fraction $x$ and transverse momentum component $k_T$. We have included this $k_T$-dependence in 
order to incorporate the Cronin effect in p+A collisions. We assume a Gaussian form in this 
variable~\cite{Owens:1986mp,Wang:1998ww},
\ben
f_{a/N}(x_a, k_{aT}^2) = f_{a/N}(x_a) \frac{1}{\pi\langle k_T^2\rangle} e^{-k_{aT}^2/\langle k_T^2\rangle},
\label{gauss}
\een
where $f_{a/N}(x_a)$ are the usual collinear PDFs in a nucleon.

On the other hand, prompt photon production in p+p collisions has two components 
- the so-called direct and fragmentation contributions~\cite{Owens:1986mp,Gordon:1993qc}:
\ben
\frac{d\sigma}{dy d^2p_T}=\frac{d\sigma^{\rm dir.}}{dy d^2p_T}+\frac{d\sigma^{\rm frag.}}{dy d^2p_T}.
\een
The fragmentation contribution is given by:
\ben
\frac{d\sigma^{\rm frag.}}{dy d^2p_T}&=&K \frac{\alpha_s^2}{s}\sum_{a,b,c}\int \frac{dx_a}{x_a}d^2k_{aT} \, 
f_{a/N}(x_a, k_{aT}^2)\int \frac{dx_b}{x_b}d^2k_{bT} \, 
f_{b/N}(x_b, k_{bT}^2) 
\nnu
&&\times
\int \frac{dz_c}{z_c^2} \, D_{\gamma/c}(z_c)H_{ab\to c}(\hat s,\hat t,
\hat u)\delta(\hat s+\hat t+\hat u).
\label{frag}
\een
In other words, it is exactly the same as Eq.~(\ref{light}) if one replaces the parton-to-hadron FF 
$D_{h/c}(z_c)$ by the parton-to-photon FF $D_{\gamma/c}(z_c)$. On the other hand, the direct 
contribution can be written as:
\ben
\frac{d\sigma^{\rm dir.}}{dy d^2p_T}=
K\frac{\alpha_{\rm em} \alpha_s}{s}\sum_{a,b} 
\int \frac{dx_a}{x_a} d^2k_{aT} f_{a/N}(x_a, k_{aT}^2) \, 
\int \frac{dx_b}{x_b}d^2k_{bT} f_{b/N}(x_b, k_{bT}^2) \, 
H_{ab\to \gamma}(\hat s,\hat t,\hat u)\delta\left(\hat s+\hat t+\hat u\right),
\label{dir}
\een
where $H_{ab\to \gamma}$ are the well-known partonic hard-scattering functions for 
direct photon production~\cite{Owens:1986mp,Xing:2012ii,Kang:2011rt}.

In the numerical calculations we choose $\langle k_T^2\rangle_{pp}=1.8$ GeV$^2$ in p+p 
collisions~\cite{Vitev:2003xu,Vitev:2002pf}. We use CTEQ6L1 PDFs \cite{Pumplin:2002vw}, 
fDSS parametrization for parton-to-hadron FFs~\cite{deFlorian:2007aj} and GRV parametrization for 
parton-to-photon FFs~\cite{Gluck:1992zx}. We fix the factorization and renormalization scales 
to the transverse momentum of the produced particle, $\mu_f=\mu_r=p_T$. 
We find that a $\mathcal{O}(1)$ $K$-factor can give a good description of hadron 
production in both RHIC and LHC energies, see Fig.~\ref{pp_comp}.  For prompt photons the
$K$-factor values are slightly larger and also $\sqrt{s}$-dependent but, importantly, the shape is
well-described. In the left panel of Fig.~\ref{pp_comp}, we compare our  calculation 
to RHIC photon and $\pi^0$ data at $\sqrt{s}=200$~GeV~\cite{Adare:2012yt, Adare:2007dg}. 
We also consider LHC photon data at $\sqrt{s}=7$~TeV~\cite{Chatrchyan:2011ue} and charged
 hadron data  at $\sqrt{s}=2.76$~TeV~\cite{CMS:2012aa}, shown in the right panel. The 
 leading order pQCD formalism  provides a reasonable baseline description of particle production 
 in p+p collisions. The precise  value of the $K$-factor is not important for the study of the nuclear 
modification   of  inclusive particle cross sections in p+A collisions (it cancels in ratios).     
We will incorporate nuclear effects relevant to  p+A reactions in this formalism in the next subsection.

\bef
\psfig{file=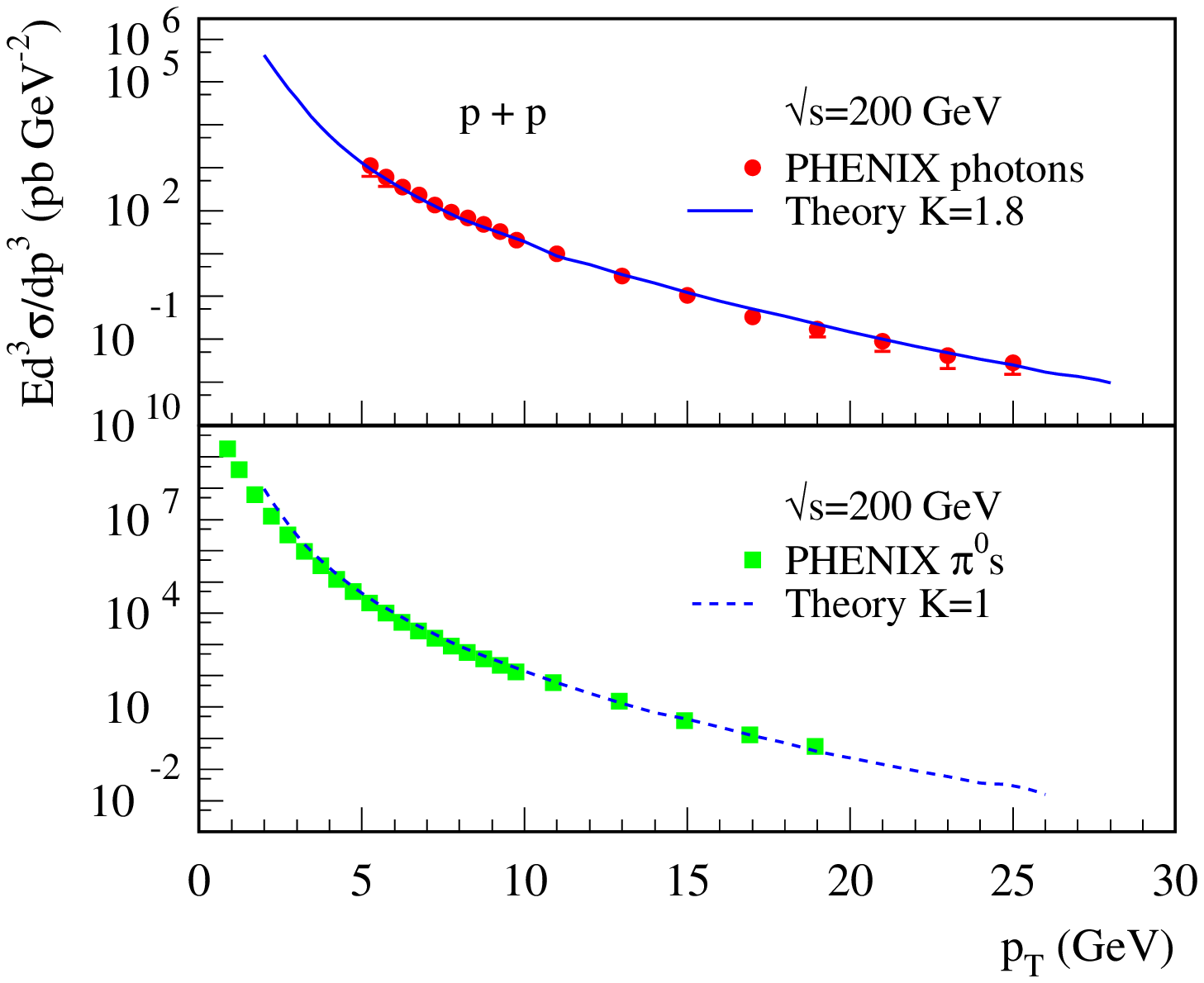, width=3.35in}
\hskip 0.3in
\psfig{file=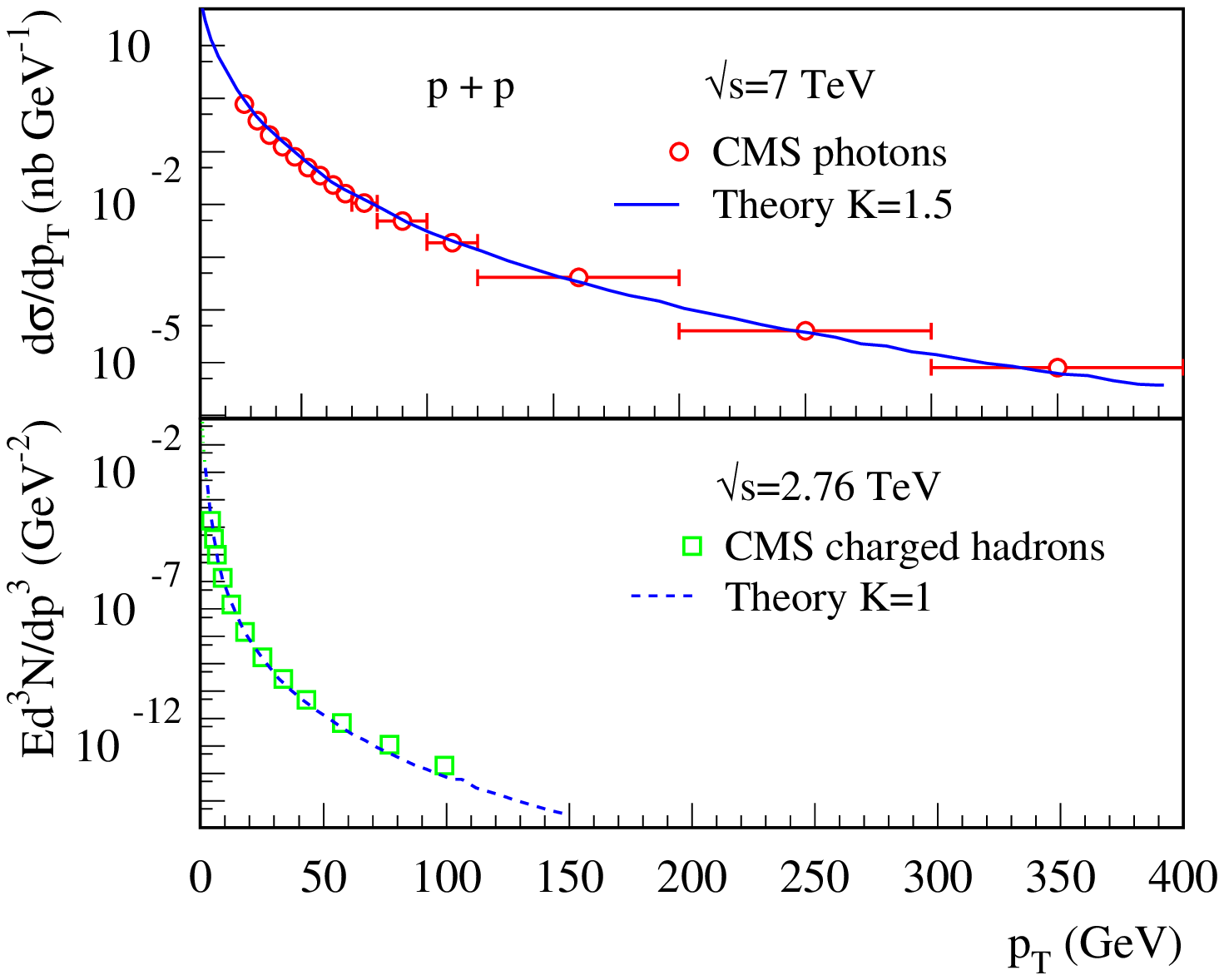, width=3.35in}
\caption{Left panel: our calculation at $\sqrt{s}=200$ GeV and rapidity $y=0$ is compared to RHIC photon data 
(top)~\cite{Adare:2012yt}  and $\pi^0$ data (bottom)~\cite{Adare:2007dg}. We choose $K=1$ for $\pi^0$ and 
$K=1.8$ for photon production. Right panel: comparison to LHC photon data at $\sqrt{s}=7$~TeV 
and $|y|<0.9$~\cite{Chatrchyan:2011ue} and charged hadron data at $\sqrt{s}=2.76$ TeV and 
$|y|<1$~\cite{CMS:2012aa}. We choose $K=1.5$ for photon and $K=1$ for charged hadron production.}
\label{pp_comp}
\eef

%%%%%%%%%%%%%%
\subsection{Cold nuclear matter effects}
The p+A  (e.g. d+Au or p+Pb) nuclear modification factor $R_{pA}$ is usually defined as:
\ben
R_{pA} = \frac{1}{\langle N_{\rm coll}\rangle}
\left.
\frac{d\sigma_{pA}}{dy d^2p_T} \right/\frac{d\sigma_{pp}}{dy d^2p_T},
\een
where $\langle N_{\rm coll}\rangle$ is the average number of binary collisions. The deviation 
of $R_{pA}$ from unity reveals the presence of CNM effects in p+A collisions. 
A  variety of CNM effects can affect particle production. In this manuscript we 
discuss the ones that have been theoretically evaluated as arising from the elastic, inelastic and coherent scattering 
of partons in large nuclei~\cite{Vitev:2006bi}. We also account for the proton and neutron composition of 
the interacting nucleus. In  particular, we implement  the isospin effect, 
Cronin effect, cold nuclear matter energy loss and dynamical shadowing. These effects have been well 
documented in the literature. We give a brief overview of their implementation in preparation 
for LHC predictions. 

{\it Isospin effect.} Isospin effect can be easily accounted for on average in the nPDFs for a 
nucleus with atomic mass $A$ and $Z$ protons via~\cite{Wang:1998ww,Kang:2008wv}:
\ben
f_{a/A}(x) = \frac{Z}{A} f_{a/p}(x) + \left(1-\frac{Z}{A}\right)f_{a/n}(x).
\label{iso}
\een
In Eq.~(\ref{iso}), $f_{a/p}(x)$ and $f_{a/n}(x)$ are the PDFs inside a proton and neutron, respectively.
The PDFs in the neutron are related to the PDFs in the proton via isospin symmetry. 

{\it Cronin effect.} Theoretical approaches to understanding the Cronin effect are well documented 
in~\cite{Accardi:2002ik}. It can be modeled via multiple initial-state scatterings of the partons 
in cold nuclei and the corresponding induced parton transverse momentum broadening~\cite{Zhang:2001ce,Ovanesyan:2011xy}.
In particular, 
if the parton distribution function $f_{b/A}(x_b, k_{b,T}^2)$ has a normalized Gaussian form, the random 
elastic scattering induces further $k_T$-broadening in the nucleus~\cite{Wang:1998ww,Vitev:2008vk}:
\ben
\langle k_{b, T}^2\rangle_{pA} = \langle k_{b, T}^2\rangle_{pp} 
+ \left\langle \frac{2\mu^2 L}{\lambda_{q,g}}\right\rangle \zeta.
\een
Here $k_{b,T}$ is the transverse momentum component for the parton prior to the hard scattering,
$\zeta=\ln(1+\delta p_T^2)$~\cite{Vitev:2003xu,Vitev:2002pf}, and we choose $\delta = 0.14$ GeV$^{-2}$, 
 $\mu^2=0.12$ GeV$^2$, $\lambda_g= C_F/C_A \lambda_q$ = 1 fm. These parameters can describe 
reasonably well RHIC data.

{\it Cold nuclear matter energy loss.} As the parton from the proton undergoes multiple scattering in 
the nucleus before the hard collisions, it can lose energy due to medium-induced gluon bremsstrahlung. 
This effect can be easily implemented as a momentum fraction shift in the PDFs
\ben
f_{q/p}(x_a) \to f_{q/p}\left(\frac{x_a}{1-\epsilon_{\rm eff.}}\right),
\quad
f_{g/p}(x_a) \to f_{g/p}\left(\frac{x_a}{1-\epsilon_{\rm eff.}}\right).
\label{eloss}
\een
Ideally,  Eq.~(\ref{eloss}) should include a convolution over the probability distribution
of cold nuclear matter energy loss~\cite{Neufeld:2010dz}. However, concurrent implementation of
such distribution together with the Cronin effect and coherent power corrections is computationally very 
demanding. The main effect of the fluctuations due to multiple gluon emission is an effective reduced
fractional energy loss $\epsilon_{\rm  eff.}$ relative to the mean value $\langle \epsilon \rangle = 
\langle   \sum_i\frac{\Delta E_i}{E}   \rangle$. Here, the sum runs over all medium-induced gluons.
In this work we follow Ref.~\cite{Sharma:2009hn} and use $ \epsilon_{\rm eff.} = 0.7 \langle \epsilon \rangle $.
The average cold nuclear matter energy loss is obtained by integrating the initial-state medium-induced 
bremsstrahlung spectrum, first  
derived in~\cite{Vitev:2007ve}. It also depends on the typical transverse momentum transfer squared
$\mu^2$  per interaction between the parton and the medium and the gluon mean free path $\lambda_g$. 
For this reason, the parameters are constrained to be the same as in the implementation of the
Cronin effect, $\mu^2=0.12$~GeV$^2$  and $\lambda_g=1$~fm in our comparison to 
RHIC data in the next section. This calculation of initial-state cold nuclear matter energy loss 
has been shown to give a good description of the nuclear modification of Drell-Yan production 
in fixed target experiments~\cite{Neufeld:2010dz}.

{\it Dynamical shadowing.} Power-suppressed resummed coherent final-sate scattering of the struck partons 
leads to shadowing effects (suppression of the cross section in the small-$x$ 
region)~\cite{Qiu:2004da}. The effect can be interpreted as a generation of dynamical parton mass
in the background gluon field of the nucleus~\cite{Qiu:2004qk}. It is included via:
\ben
x_b\to x_b \left(1+C_d \frac{\xi^2(A^{1/3}-1)}{-\hat t}\right),
\een
where $x_b$ is the parton momentum fraction inside the target nucleus, $C_d=C_F (C_A)$ if the 
parton $d=q (g)$ in the partonic scattering $ab\to cd$. $\xi^2$ represents a characteristic 
scale of the multiple scattering per nucleon.  In the RHIC energy range $\sqrt{s}=200$ GeV, 
$\xi^2_q = C_F/C_A \xi^2_g = 0.12$~GeV$^2$~\cite{Qiu:2003vd,Qiu:2004da}
can give a good description for nuclear modification in d+Au collisions for both single 
hadron and di-hadron  production~\cite{Qiu:2004da,Kang:2011bp}.

\section{Theoretical model predictions in p+A collisions}

\begin{figure}[!t]
\psfig{file=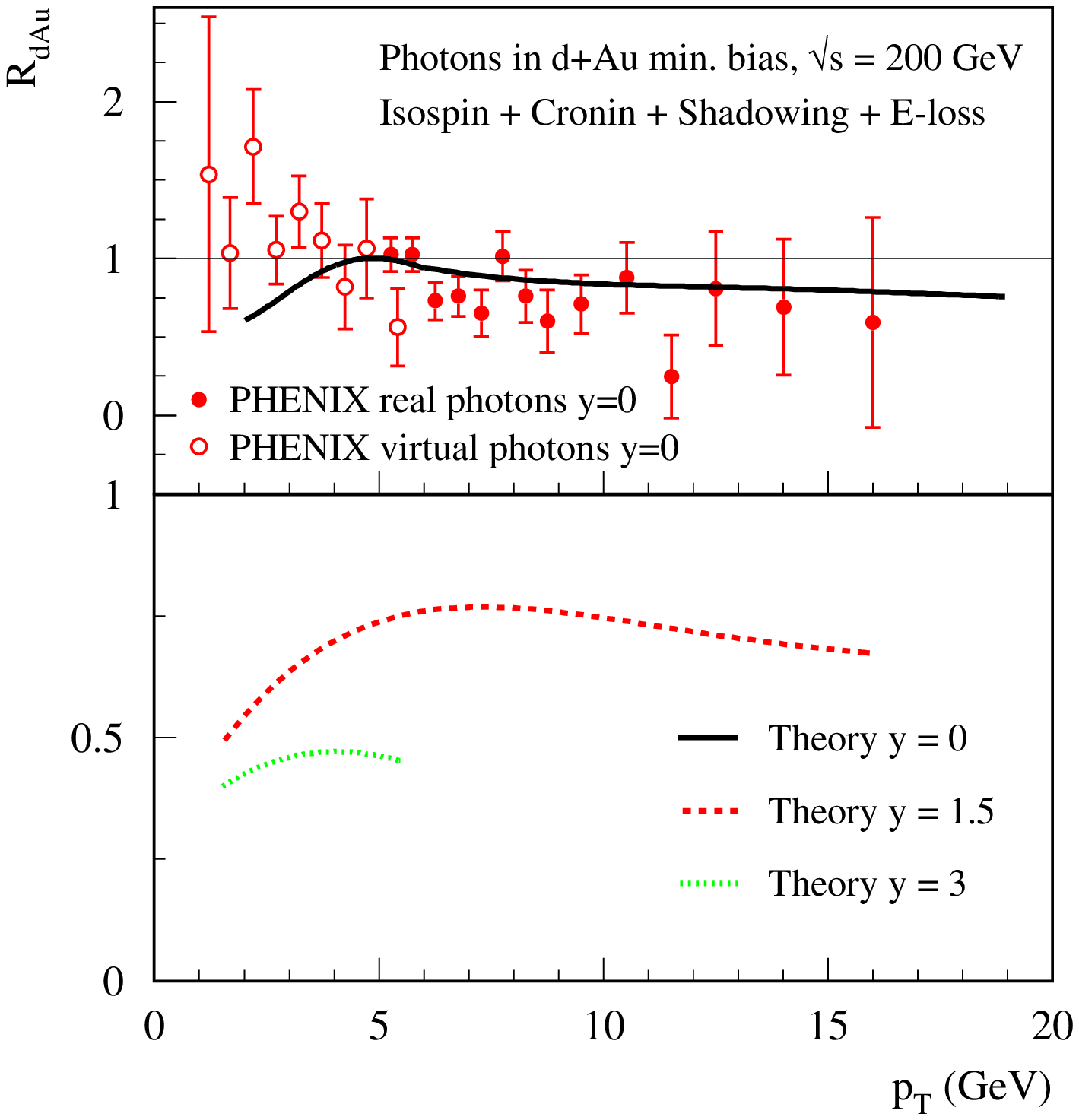, width=3.35in}
\hskip 0.3in
\psfig{file=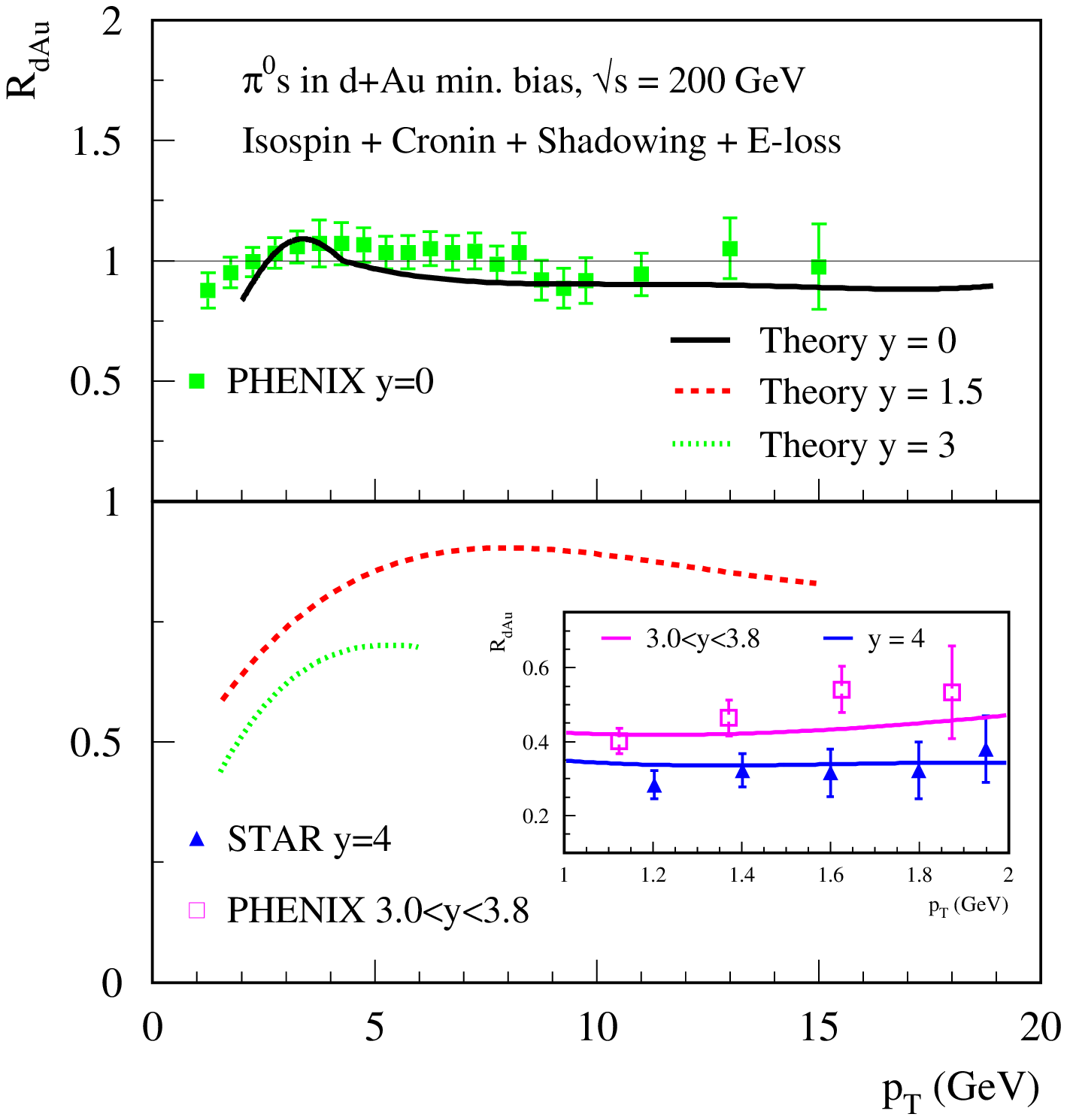, width=3.35in}
\caption{The nuclear modification factor $R_{dAu}$ is plotted as a function of transverse 
momentum $p_T$ for photons (left panel) and $\pi^0$s (right panel) at RHIC energy of $\sqrt{s}=200$ GeV.
Empty and filled circles are low-mss virtual photon and real photon experimental data 
at mid-rapidity from PHENIX~\cite{Adare:2012yt}. Filled squares are the minimum bias mid-rapidity $\pi^0$ 
data~\cite{Adler:2006wg},  filled triangles are the forward rapidity $y=4$ STAR $\pi^0$ data~\cite{Adams:2006uz}, and
empty squares is the forward rapidity $3<y<3.8$ PHENIX $\pi^0$s~\cite{Adare:2011sc}. We have also 
presented predictions for rapidity $y=1.5$ (dashed curves) and $y=3$ (dotted curves) for both photon and hadron production.}
\label{rhic}
\end{figure}

In this section we compare our theoretical model predictions to existing light hadron and direct 
photon  data in minimum bias d+Au reactions at  RHIC at
$\sqrt{s}= 200$~GeV. We also present predictions for these inclusive final states  
in minimum bias p+Pb collisions at the LHC  at $\sqrt{s}= 5$~TeV. It should be understood that in nucleus 
collisions we quote the center-of-mass energy per nucleon pair.

%%%%%%%%%%%%%%
\subsection{Nuclear modification factor in d+Au reactions at RHIC}

We have incorporated the cold nuclear matter effects discussed in the previous section 
into the leading order pQCD photon and hadron production formalism, Eqs.~(\ref{light}) and (\ref{frag}),(\ref{dir}).
Such theoretical approach is able to give a good description of single inclusive particle production in d+Au 
collisions at RHIC energies. While such calculations have been presented in the past, in this manuscript we extend
our results to direct photons at forward rapidity.  In Fig.~\ref{rhic}, we plot the minimum bias nuclear 
modification factor $R_{dAu}$  as a function of transverse momentum for both photon (left panel) and $\pi^0$ 
(right panel) production at $\sqrt{s}=200$ GeV. The photon data is from the PHENIX collaboration 
at rapidity $y=0$~\cite{Adare:2012yt}. For $\pi^0$s we have included both mid-rapidity $y=0$~\cite{Adler:2006wg} 
and forward rapidity $3<y<3.8$~\cite{Adare:2011sc} PHENIX data, and forward rapidity 
$y=4$ STAR data~\cite{Adams:2006uz}. We conclude that implementation of cold nuclear matter effects 
based on the physics of multiple parton scattering in dense QCD matter can describe both photon and hadron data quite 
well at  mid and forward rapidities. With the future RHIC upgrade program in mind, we also show predictions for 
the nuclear modification factor $R_{dAu}$ for photon and $\pi^0$ production in different kinematic regions: the dashed 
curves are for $y=1.5$ and the dotted curves are for $y=3$.

%%%%%%%%%%%%%%
\subsection{Predictions for the p+Pb run at the LHC}
We are now in good position to present predictions for  minimum bias p+Pb reactions at the LHC energy 
of $\sqrt{s}=5$~TeV.  These results will soon be confronted by new experimental data. 
The main parameters in our approach include:  the Cronin momentum broadening parameters $\mu^2, 
\lambda_{q,g} $, 
which also determine the cold nuclear matter energy loss, and the dynamical 
shadowing parameter $\xi^2$. As emphasized in~\cite{Xing:2012ii}, $\mu^2/\lambda_{q,g}$ and $\xi^2$ represent 
the strength 
of the multiple scattering between the incoming and outgoing partons and the target nucleus, thus 
proportional to the number of the soft gluons in the nuclear medium. Following~\cite{Xing:2012ii}, 
beyond the possibility of fixed $\xi^2$, which we take as the lower limit for this parameter,  we study a scenario 
where $\xi^2 \propto \sigma_{in}$, the inelastic  nucleon-nucleon scattering cross section. 
Taking into account that  $\sigma_{in}=42$ mb~\cite{Miller:2007ri}  at RHIC energy of $\sqrt{s}=200$~GeV 
and  $\sigma_{in}=70$ mb~\cite{d'Enterria:2003qs}  at the LHC energy $\sqrt{s}=5$ TeV, we consider a potential 
increase in the $\xi^2$ parameter by 67\%.  Specifically, for the upper limit  of this parameter we take
 $\xi^2 =0.20$~GeV$^2$ in our prediction for the LHC p+Pb run at $\sqrt{s}=5$ TeV. 
We will also consider a correlated enhancement in $\mu^2/\lambda_g$.

\bef
\psfig{file=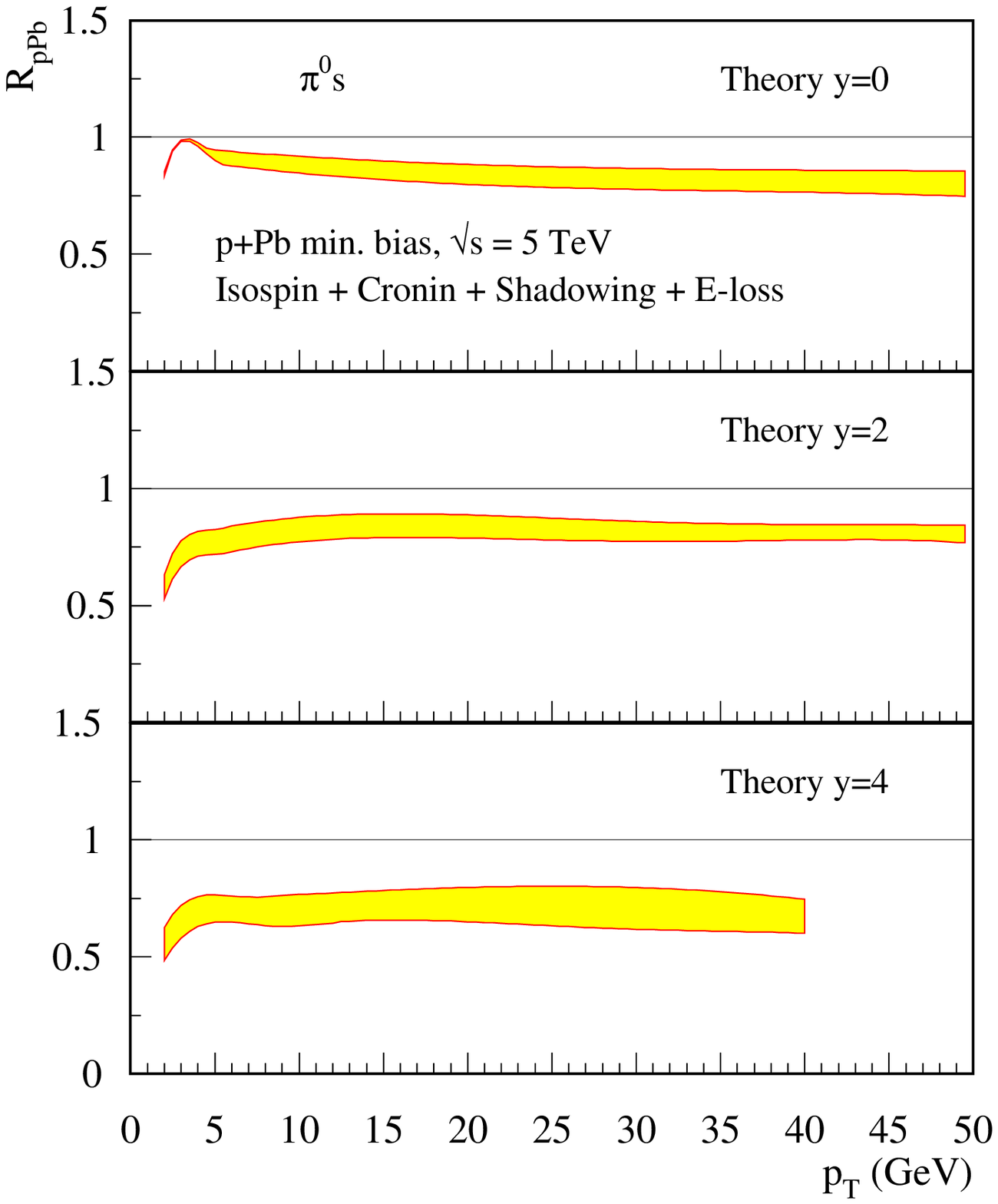, width=3.3in}
\hskip 0.2in
\psfig{file=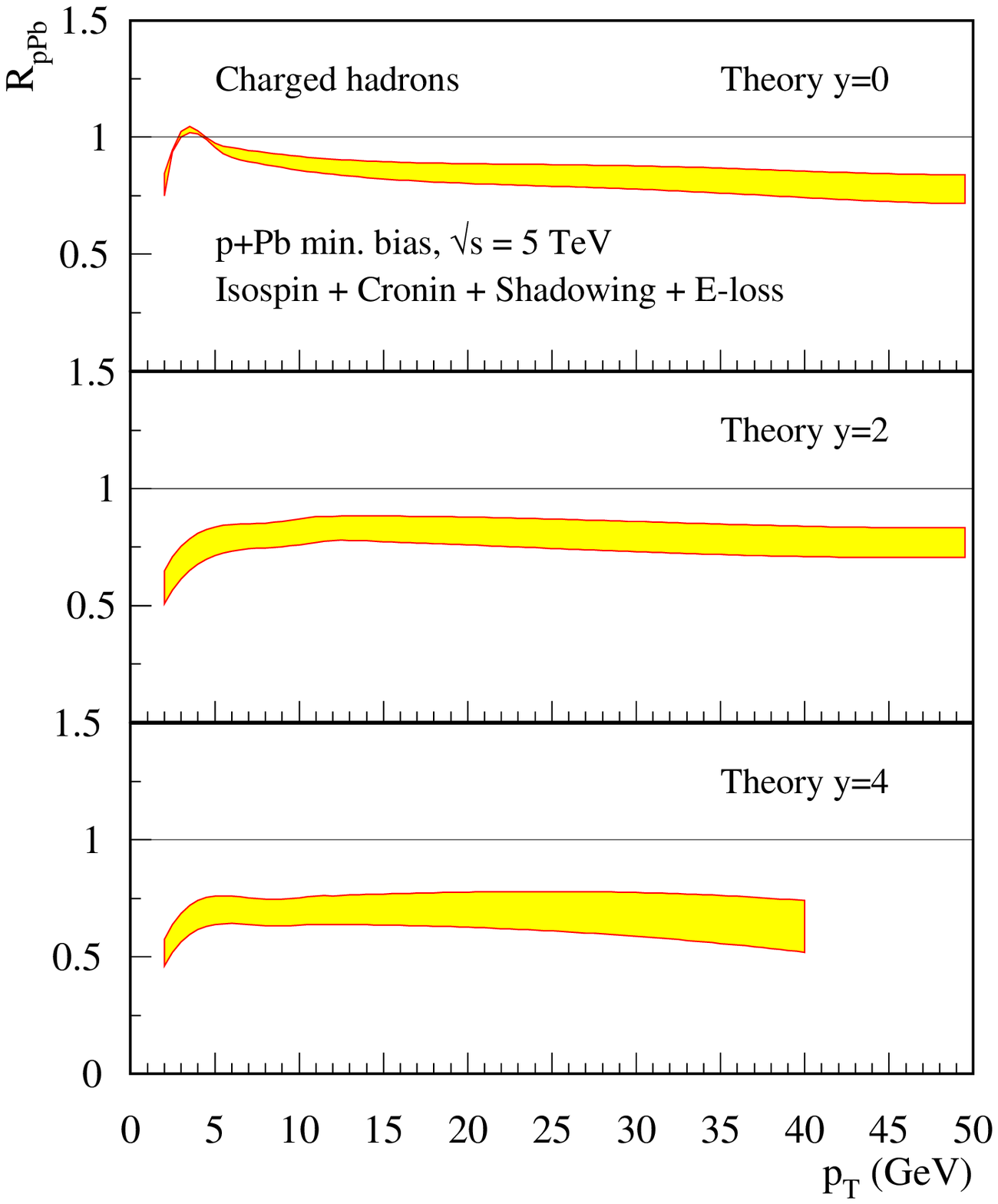, width=3.3in}
\caption{Predictions for the nuclear modification factor $R_{pPb}$ as a function of 
transverse momentum $p_T$ for $\pi^0$ (left panel) and charged hadron (right panel) production in minimum 
bias p+Pb collisions at the LHC energy of $\sqrt{s}=5$~TeV. Results  for three rapidities $y=0$, $y=2$, 
and $y=4$ are shown.}
\label{lhc_h}
\eef

\bef
\psfig{file=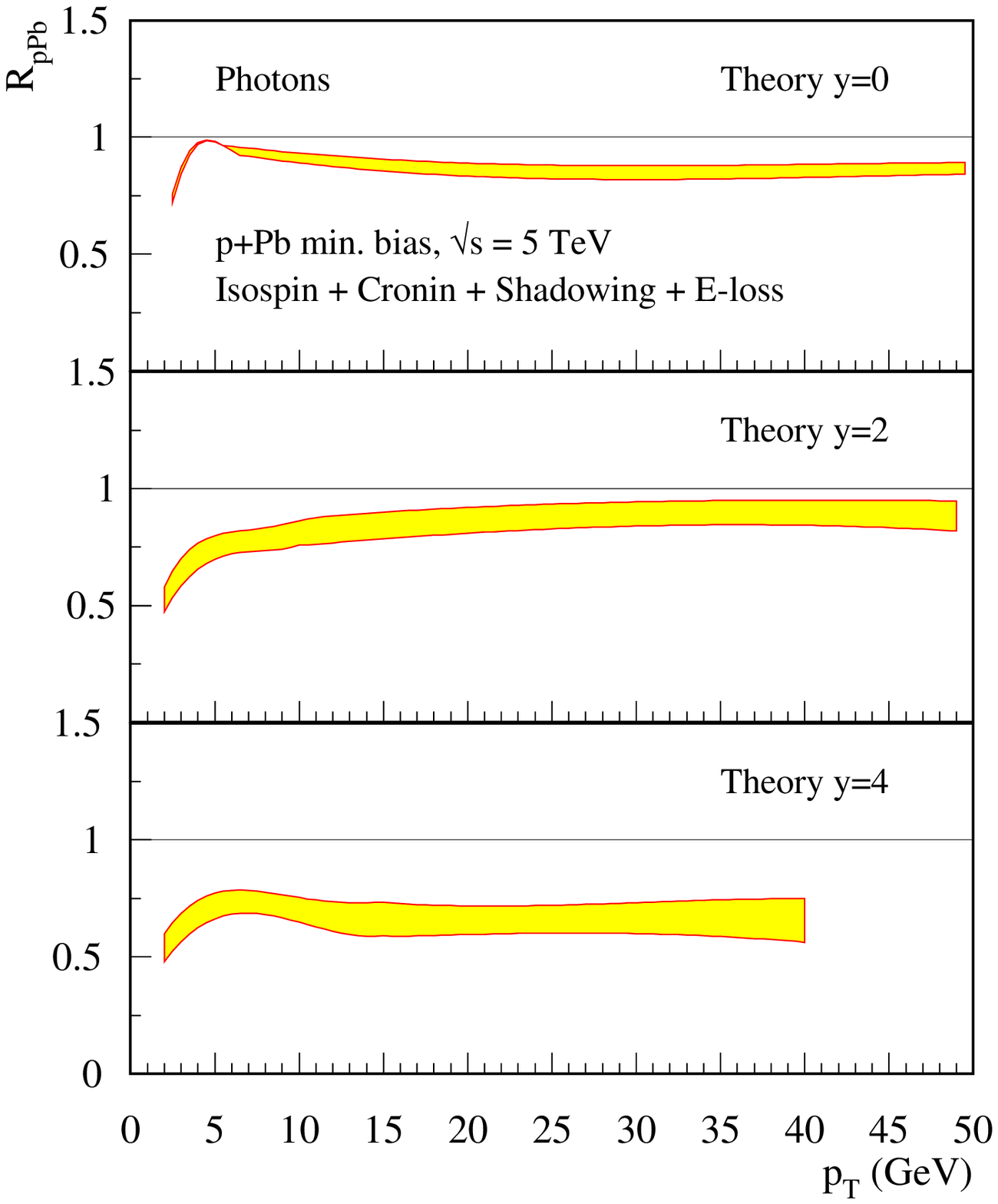, width=3.3in}
\caption{Predictions for the nuclear modification factor $R_{pPb}$ as a function of 
transverse momentum $p_T$ for photon  production in minimum 
bias p+Pb collisions at the LHC energy of $\sqrt{s}=5$~TeV   for rapidities $y=0$, $y=2$, 
and $y=4$.}
\label{lhc_phot}
\eef

In Fig.~\ref{lhc_h} we present our theoretical model predictions for the nuclear modification 
factor $R_{pPb}$ as a function of  $p_T$ for neutral pion (left panel) 
and charged hadron (right panel) production in minimum bias p+Pb collisions at the LHC energy 
of $\sqrt{s}=5$ TeV.  Rapidities $y=0$ (top), $y=2$ (middle), and $y=4$ (bottom) were considered. 
The upper edge of the bands corresponds to the RHIC parameters $(\mu^2, \xi^2)=(0.12, 0.12)$ GeV$^2$. 
The lower edge corresponds to  a potential enhancement of these parameters as discussed above, 
$(\mu^2, \xi^2)=(0.2, 0.2)$~GeV$^2$, to be tested against the experimental data.  
In Fig.~\ref{lhc_phot}, we present similar results for prompt photon production. 

The general features of these nuclear modification factors are very similar to the ones observed at RHIC 
energies. At mid-rapidity $y=0$, there is a very small ``Cronin peak'' in the low  $p_T$ region. The peak is 
very close to unity and not as pronounced as the one observed in low energy fixed target experiments.
This is because the dynamical shadowing becomes important and strongly suppress the particle production 
in this region. Furthermore, initial-state energy loss is larger due to the bigger contribution of 
diagrams with incoming gluons to the cross section. At high $p_T$ we still have  $\sim 15\%$ suppression, 
which is also due to  the cold nuclear matter energy loss. In general, at forward rapidity all CNM effects are
amplified due to the larger values of the momentum fraction $x_a$ from the incoming proton (relevant to
cold nuclear matter energy loss), the smaller values of the momentum fraction $x_b$ of the incoming nucleus
(relevant to dynamical shadowing) and the steeper falling spectra (relevant to the Cronin effect).  
As a result,  at forward rapidity and at low $p_T$ the dynamic shadowing 
effect can be most important and lead to stronger suppression of inclusive particle production while at 
high $p_T$ the suppression  is a combined effect of cold nuclear matter energy loss and the Cronin effect. Finally, the 
behavior of the $R_{pPb}$ for $\pi^0$s, charged hadrons and direct photons is qualitatively the same 
and the quantitative differences are minor. 

It is interesting to point out that our predictions for inclusive particle suppression at forward rapidities 
are for observable effects that are stronger than the effects found in
calculations using EPS09 nuclear PDFs.  On the other hand, they lead to weaker suppression 
than the CGC predictions from a hybrid formalism  in rc-BK-MC approach~\cite{Albacete:2012xq,Arleo:2011gc}. 
In other words, our predictions fall between EPS09 and CGC based approaches.

\section{Summary}

We studied the nuclear modification of high transverse momentum particle production in minimum bias
d+Au and p+Pb  collisions at  RHIC and LHC, respectively. Such nuclear modification is  manifestation 
of nontrivial and not yet fully understood QCD dynamics in large nuclei. It is, therefore, critical 
to further our understanding of high energy nuclear reactions through concurrent theoretical advances 
and comparison of model predictions to precise experimental data.
With this motivation, we presented results  from a theoretical approach that 
combines the leading order perturbative QCD baseline calculation of inclusive light hadron and photon
production with cold nuclear matter effects that arise from the elastic, inelastic and coherent
parton scattering in large nuclei. Our numerical simulations included the isospin effect, 
Cronin effect, cold nuclear matter energy loss and resummed QCD power corrections to the leading
twist results. We found that this approach can describe quite well the nuclear modification factor in 
$\sqrt{s}=200$~GeV  d+Au collisions at the RHIC for both photon and light hadron production, 
including  mid and forward rapidities. Detailed theoretical model predictions were then presented 
for inclusive particle production relevant to the recently completed p+Pb run at $\sqrt{s}=5$~TeV  at the LHC.
Our results for the nuclear modification of light hadrons and photons at forward rapidities fall 
between the ones based on EPS09  shadowing parameterization and CGC approaches. These predictions will 
soon be confronted by new experimental data to help constrain the magnitude and clarify the origin of 
cold nuclear matter effects. Within the framework of our approach, such comparison will also shed 
light on  the transport properties of cold nuclear matter.  

\section*{Acknowledgments}
This research is supported by the US Department of Energy, Office of Science, under Contract 
No.~DE-AC52-06NA25396, by the LDRD program at LANL and by the NSFC of China under Project No. 10825523.

%%%%%%%%%%%%%%

\end{document}